# Characteristic spectral features of the polarized fluorescence of human breast cancer in the wavelet domain

**Anita H. Gharekhan**

*C.U.Shah Science College, Gujarat University, Ahmedabad-380 009, India*

**Nrusingh C. Biswal**

*Electrical and Computer Engineering Department, University of Connecticut, CT 06269-2157, USA*

**Sharad Gupta**

*Department of Biomedical Engineering, Tufts University, Medford, MA 02155, USA*

**Prasanta K. Panigrahi**

*Indian Institute of Science Education and Research (IISER)-Kolkata, BCKV Main Campus, Mohanpur, Nadia-741252, India*

and

**Asima Pradhan**

*Department of Physics and Centre for Laser Technology, Indian Institute of Technology, Kanpur, India*

**Corresponding author: asima@iitk.ac.in**

Wavelet transform of polarized fluorescence spectra of human breast tissues is found to localize spectral features that can reliably differentiate normal and malignant tissue types. The intensity differences of parallel and perpendicularly polarized fluorescence spectra are subjected to investigation, since the same is relatively free of the diffusive background. A number of parameters, capturing spectral variations and subtle changes in the diseased tissues in the visible wavelength regime, are clearly identifiable in the wavelet domain. These manifest both in the average low pass and high frequency high pass wavelet coefficients.

Keywords: Florescence Spectroscopy; Wavelet Transform; Biomedical Optics; Breast Cancer

**Introduction**

Breast cancer has emerged as the most common disease amongst women. Although the risk factor for Asian women has been estimated to be one-fifth to one-tenth that of women in North America and Western Europe, it still is the second most malignant condition.[1,2] Apart from genetic predisposition, a number of factors like diet, exercise, environment, etc., are being recognized to play major roles in the growth of the disease.[3] Early diagnosis is still not possible through conventional diagnostic techniques. If diagnosed early, breast cancer is also one of the most treatable forms of cancer. The requirement of continuous monitoring for breast malignancy of a significant percentage of women population has led to an intense search for safe, reliable and fast diagnostic methods.

Optical diagnosis methods are now emerging as viable tools for tumor detection. Of these, fluorescence techniques are being increasingly employed to investigate both morphological and biochemical changes in different tissue types, for eventual application in the detection of tumors at an early stage.[4-6] Fluorescence spectroscopy is well suited for the diagnosis of cancerous tissues because of its sensitivity to minute variations in the amount and the local environment of the native fluorophore present in the tissues.[7-12] A number of fluorophores, ranging from structural proteins to various enzymes and coenzymes, some of which participate in the cellular oxidation-reduction processes, are present in the human tissue and can be excited by ultraviolet and visible light.[7] The fluorophore, FAD (Flavin Adenine Dinucleotide), its derivatives and porphyrin are particularly useful as fluorescent markers, since they fluoresce in the higher wavelength visible region, when excited by lower wavelength visible light, thereby avoiding the potentially harmful ultraviolet radiation.

The fluorescence emission can differ significantly in normal and cancerous tissues due to the differences in concentrations of absorbers[13,14] and scatterers, as also the scatterer sizes.[15] Some of the fluorophores like porphyrin give rise to weak fluorescence, since accumulation of porphyrin in tissue is a time consuming process. The absorption in the visible range occurs primarily due to the presence of blood, whose amounts vary in various tissue types.[16] The presence of scatterers leads to randomization of light, thereby generating a depolarized component in the fluorescence spectra. Polarized fluorescence spectroscopy is useful in isolating the characteristic spectral features from the diffuse background. The parallel component of the fluorescence suffers fewer scattering events. In comparison, the intensity of the perpendicular component is not only affected more by scatterers, but is also quite sensitive to absorption, since the path traversed by the same in the tissue medium is more. Hence, the difference of parallel and perpendicular intensities, apart from being relatively free from the diffusive component,[17,18] can be quite sensitive to microscopic biochemical changes including the effects of absorption in different tissue types.

The analysis of spectral data involve both physical[19-22] and statistical[16,23] modeling of tissue types, as also statistical methods, e.g., principal component analysis[16,24] for extracting distinguishing parameters for diagnostic purposes. The fact that biological tissues are complex systems, possessing substantial variations among individual patients, depending upon various factors such as age, progress of the disease, etc., makes modeling of the same rather difficult.[25] In using statistical tools, difficulty often arises in relating the statistically significant quantities to physically transparent spectral variables.

In recent times, wavelet transform has emerged as a powerful tool for the analysis of transient data and is particularly useful in disentangling characteristic variations at different scales.[26] This linear transform isolates local features and leads to a convenient dimensional reduction of the data in the form of average coefficients, resembling the data itself. The wavelet coefficients, at various levels, encapsulate the variations at corresponding scales. Earlier studies, of the perpendicular component of the fluorescence spectra, by some of the present authors have indicated the usefulness of wavelet transform in identifying characteristic spectral features in the high frequency spectral fluctuations.[27,28] The statistical properties of these fluctuations showed differences between the tissue types.

In this paper, we first present the results of a systematic analysis of both the average and high-pass coefficients of the wavelet transform, of the difference between parallel and perpendicular components of the fluorescence spectra, from normal and malignant human breast tissues. The average coefficients in the wavelet domain are less sensitive to experimental and statistical uncertainties. A number of parameters, capturing spectral features and subtle changes in the intensity profile of the diseased tissues, as compared to their normal counterparts, are identified in the wavelet domain. The physical origin of one of the distinguishing parameters may be ascribed to the changes in the concentration of porphyrin and the density of cellular organelles present in tumors.[15,29] We then studied the behavior of the high pass coefficients and find distinguishing features between tissue types.

**Wavelet Transform: A brief overview**

Wavelet transform is known as a mathematical microscope, which isolates the variation and averages systematically at different scale due to its multi resolution ability, hence one can simultaneously have both time and frequency information, which is not possible through standard approaches like Fourier transform. The data under consideration is separated out into high frequency and low frequency components at multiple scales, known respectively as high pass and low pass coefficients. For example, high pass coefficients at level-1 represent variations at smallest scale and the subsequent higher level coefficients represent variations over bigger window sizes. The low pass coefficients at various levels represent average behavior of the data over corresponding window sizes. In discrete wavelet transform (DWT), the basis functions consist of father wavelet $\varphi(x)$ and mother wavelet $\psi(x)$ satisfying

$$\qquad , \quad \text{and} \qquad (1)$$

$\varphi$ and $\psi$ are also square integrable:

A completely orthonormal basis is constructed through translated and scaled versions of $\varphi(x)$ and $\psi(x)$. Explicitly, $\varphi(x - k) \equiv \varphi_k(x)$ and $\psi_{j,k} \equiv 2^{j/2} \psi(2^j x - k)$, for $-\infty \leq k \leq \infty$ and $0 \leq j \leq \infty$, provide a complete orthonormal basis through which any finite energy signal $f(t) \in L^2(R)$ [30] can be expanded as

Here $c_k$'s are the low pass and $d_{j,k}$'s are the high pass coefficients. For our analysis, we will make use of Haar wavelets, in which $c_k$'s can be interpreted as average of data points of appropriate size, around space location $k$, $d_{j,k}$'s are the high pass coefficients at level $j$. These represent differences around location $k$, over a window size which depends on level $j$. Explicitly the wavelet coefficients are given by

For our application, we make use of the simplest Haar wavelet, since the interpretation of the wavelet coefficients is quite transparent here; it is also free from artifacts arising due to the finite size of the data. Haar basis is special, since it is symmetric and compactly supported. Below we explicate the procedure of extracting high-pass and low-pass coefficients for Haar wavelets, which makes their physical meaning clear. As mentioned before, discrete wavelet transform carries out a local weighted averaging and differentiation operation on a given, uniformly sampled data set. In case of Haar wavelets, this operation, modulo normalization, is precisely arithmetic averaging and differentiation. For illustration, let us consider a data set having four data points. At the lowest scale, i.e., level-1, the low pass (average) and high pass (wavelet) coefficients for the data set comprising of four data points a, b, c, d are respectively given by,

Because of Parseval's theorem, the energy/power i,e., is conserved:

This explains the origin of the normalizing factor in front of the coefficient. The fact that, there are four data points, explains the presence of four independent coefficients, two low-pass and two high-pass, from which, the data points can be reconstructed by the algebraic operations of addition and subtraction. One observes the absence of coefficients like and etc; the presence of which would have led to over counting. A straight-forward binning procedure with window size two would have yielded the low-pass coefficients.

At the second level, the window size is doubled, which leads to averaging and differentiation at a bigger scale involving four points, to produce the level -2, low-pass and high pass coefficients:

In general, if the data length is , one can carry out an N-level decomposition. One notices that the window size is progressively increased in units of two. In the first level decomposition one obtains low-pass and high-pass coefficients, each of which are exactly half of the data size. In the next level, the level one low-pass coefficients are decomposed into level two low-pass coefficients and level two high-pass coefficients, each of size half that of the level one low-pass coefficients. This procedure is then carried on to the desired level of decomposition. This dyadic decomposition arises due to orthogonality and completeness of the Haar basis. Much like the orthogonal components of a vector along x, y and z directions, the obtained wavelet and low-pass coefficients are completely independent of each other. Instead of Haar, one can make use of wavelets like Daubechies basis, to isolate polynomial trends of the spectra from fluctuations. Lorentzian wavelets can be used to extract average Lorentzian line shapes from fluctuations, in spectral data analysis.

The pictorial demonstration of discrete wavelet transform is presented in Fig.1 where N=0

represents the original signal. It is worth mentioning that in case of Haar wavelet finding low pass coefficients at different levels amounts to binning of data in appropriate window sizes.

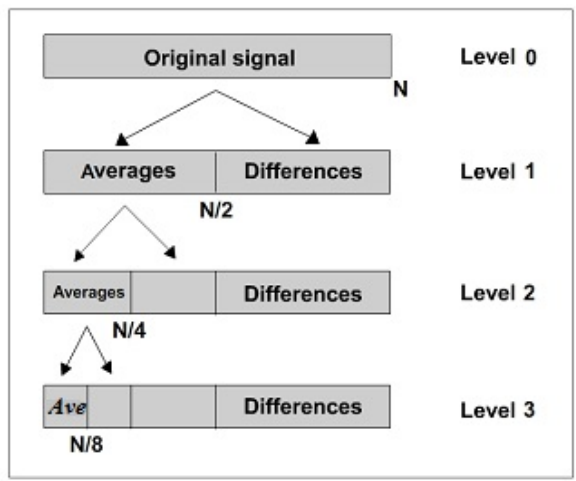

Fig.1. Pictorial demonstration of Discrete Wavelet Transform

**Experimental procedure**

The samples were excited by 488-nm wavelength plane polarized light from an Ar-ion laser (Spectra Physics 165, 5 W). The unpolarized and polarized fluorescence spectra were collected in right angle geometry using triplemate monochromator (SPEX-1877E and PMT, RCA C-31034). For polarized fluorescence, a depolarizer was used after the analyzer, in order to ensure that there was no preference of the selected directions of polarized fluorescence by the detection system. The components of fluorescence light that are parallel and perpendicular to the incident polarized light were measured in the 500- to 700-nm wavelength region. The tissue samples were kept in moist saline and frozen (at $4^0C$) immediately after biopsy. After the biopsy, a part of the tissue sample was sent for histopathology and the other part was used for fluorescence measurements. The experiment was performed within a few hours of the surgery after thawing the sample.

We have considered 192 intensity values, from 500nm to 691nm for analysis. The level-1 low pass coefficients are the averages of nearest neighbour data points; level-2 low pass coefficients represent averages of 4-data points etc. It is clear that, there are 96 level-1 coefficients, 48 level-2 coefficients etc. For 192 data points, we can have a six level decomposition, with 3 sixth level low pass coefficients. It is observed that, progressive averaging of data in discrete wavelet transform reduces statistical and experimental uncertainties. The variations at lower level represent the high frequency components, corresponding to variations in smaller window sizes. As is characteristic of wavelet transform, it is seen that the normalized low pass coefficients resemble the data. While computing wavelet power one takes the sum of the square of wavelet coefficients at particular level e.g., low pass power at level-4 corresponds to sum of square of low pass coefficients of $4^{th}$ level. For the purpose of normalization, we divide the power of a given level by the total power P:

which is the sum of the squares of all the data points. Similarly one can compute the normalized high pass power.

In our earlier studies,[27,28] we have investigated the statistical properties of the lower level high pass coefficients and have observed significant differences between cancer and normal tissues. In the present paper, the structure of the low pass coefficients and low pass power of six levels will be investigated, as also statistical behavior of the low level high pass coefficients.

**Results and Discussion**

In total, 28 breast cancer tissue samples were studied, out of these, 23 samples came with their normal counterparts. The tissue samples were excited by 488 nm wavelength polarized light and the parallel and perpendicularly polarized fluorescence light were measured from 500 to 700 nm. Differences of parallel and perpendicular components of fluorescence intensity ($I_\parallel - I_\perp$) versus wavelength profiles for all the tissue samples were analyzed by Haar wavelets,[30] for which purpose we have considered the first 192 intensity values. Typical samples of the spectra are seen in Fig. 2 and 3.

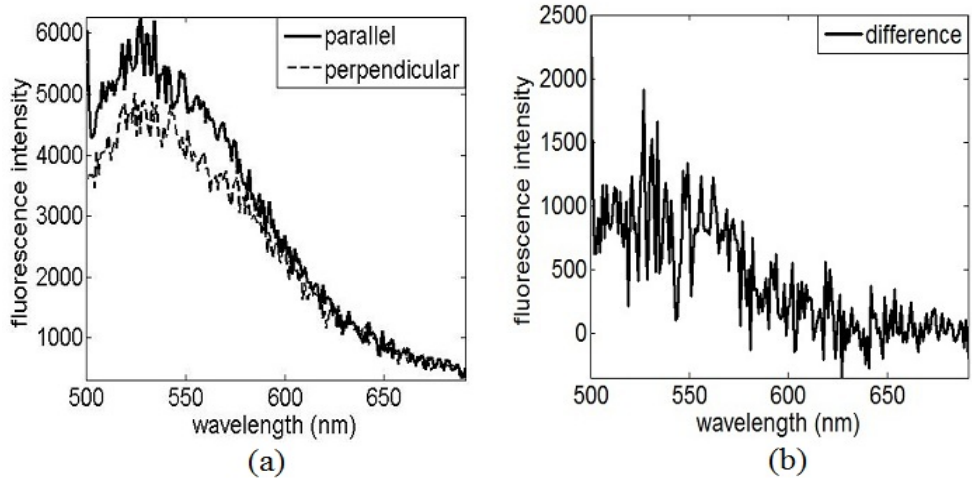

Fig.2. Typical plots of (a) parallel and perpendicular components (b) difference of the parallel and perpendicular components of intensities of fluorescence spectra of normal tissue.

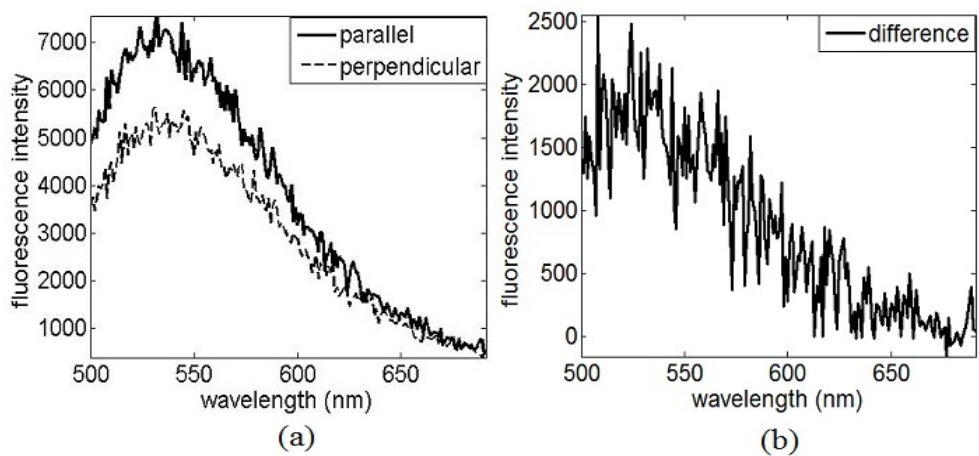

Fig.3. Typical plots of (a) parallel and perpendicular components (b) difference of the parallel and perpendicular components of intensities of fluorescence spectra of cancer tissue.

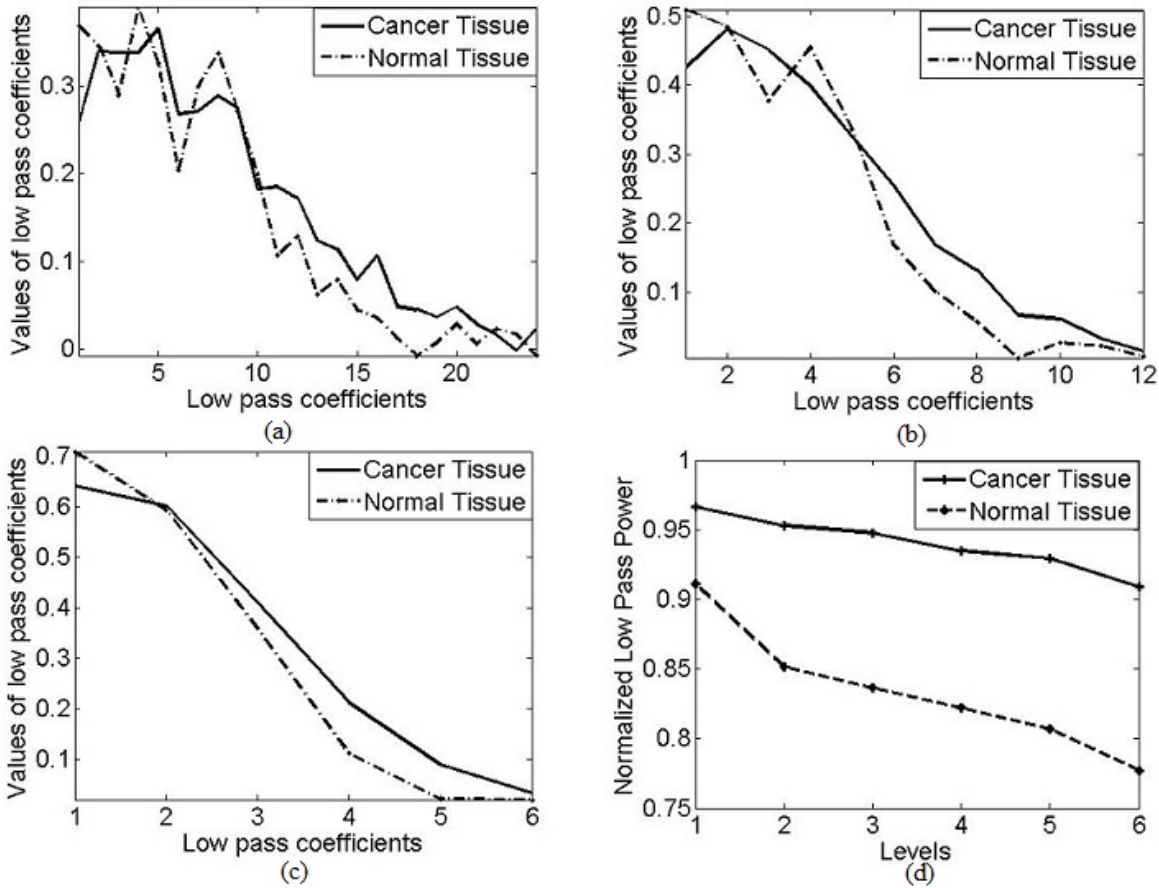

Fig.4. Plots of low pass coefficients of cancer and normal tissues for (a) level-3 (b) level-4 and (c) level-5 average coefficients. The fourth-level coefficients highlight the 630nm weak emission peak in cancerous tissue. 4(d) shows a typical low pass power plot of the difference of parallel and perpendicular components of the fluorescence spectra of cancer and normal tissues. The rate of decrease of low pass power, as a function of levels, is slower in cancer tissues.

It is found that a clear difference between cancer and normal tissues emerge at fourth level. In particular, relatively higher value for a normalized low pass coefficient at the third quarter of the 4th level decomposition is seen in cancer tissues as shown in Fig. 4(b). It should be noted that at the 4th level, each coefficient corresponds to average of 16 original intensity values. This structure is absent at the fifth level, as seen in Fig. 4(c), because of averaging over a bigger window size and is not transparent at the third level as is clear from Fig. 4(a). So a suitable block averaging highlights the presence of a weak fluorophore (porphyrin) in cancerous spectra which is not present in the normal case. The 5$^{th}$ level normalized low pass coefficients do not show significant variation between cancer and normal tissues, except at the tail end, where the cancer values fall off slowly as compared to their normal counter parts. For the purpose of comparison between normal and cancer tissue fluorescence the normalized low pass coefficients at 3$^{rd}$, 4$^{th}$ and 5$^{th}$ levels are shown in Fig. 4. As pointed out earlier, at the 4$^{th}$ level the normalized low pass coefficients of the cancer tissue fluorescence show a minor peak at the 3$^{rd}$ quadrant. The same is averaged out at the 5$^{th}$ level. At 3$^{rd}$ level, one sees many minor peaks apart from one at the 3$^{rd}$ quadrant. Significantly the normalized wavelet powers representing the strength of the variation at different levels also differ between normal and cancer tissues at the same level. A typical plot revealing this aspect is given in Fig. 4(d). It may be noted from equation(5) that the low pass power is the sum of square of low pass coefficients at that level and hence would highlight the differences in normalized low pass coefficients between normal and cancerous tissues. The behavior of the normalized low pass powers at different levels also reveal substantial differences between tissue types, as is clearly indicated in Table-1.

In case of Haar wavelet fourth level corresponds to averaging of 16 intensity values of original data set. In third level it corresponds to averaging 8 points whereas in fifth level it corresponds to averaging of 32 points. The fact that porphyrin peak manifests at the fourth level, indicates that the spectral width of this fluorophores is of the order of the above window size. The absence of this fluorophore in case of normal samples makes the intensities fall faster in 620-670nm regime. We would like to add that we have done a block averaging with window size ranging from 10, 12, 14, 16 and 18 nm. It is found that the peak at the third quadrant begins to manifest from the window size 14 nm and starts decreasing around 18 nm.

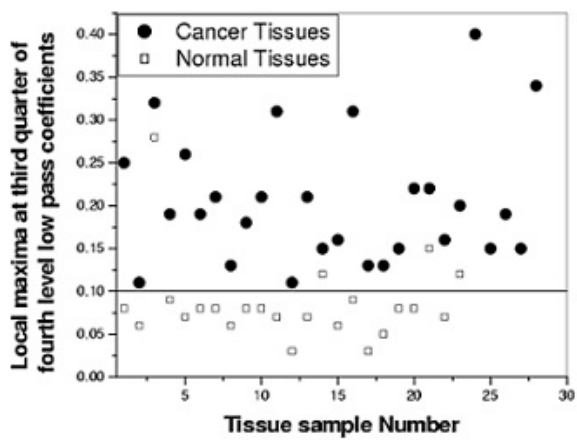

Fig.5. Local maxima at third quarter of fourth level low-pass coefficients of cancer and normal breast tissues.

The normalized low-pass coefficients in the third quarter of the 4th level have been considered for illustrating the minor peak. The local maxima at third quarter of fourth level normalized low-pass coefficients of cancer samples are more than 0.1 while those of normal tissues are less than 0.1, with a sensitivity of 100 % and specificity of 83 %. The scatter plot is shown in Fig.5. It should be noted here that the values for normal tissues which are more than 0.1 still show lower values than the corresponding tumors, consistent with all the other samples. Thus intra-patient diagnosis gives a clear distinction between cancer and normal tissues. Variations in inter-patient diagnosis may be due to the fact that, the growth of tumor depends on genetic (major genes, modifier genes) and non-genetic factors (birth, age, weight/diet, exercise, environmental exposures, etc).[3] The above fourth level normalized low-pass coefficient, originate from the fluorescence signals around 630nm of the original data, corresponding to the porphyrin emission peak.

FAD and porphyrin are the major fluorophores that fluoresce in the visible wavelength regime, with peak intensities at 530 and 630 nm respectively. These fluorophores are considered as contrast agents for cancer detection.[7,11] It has been suggested that deficiency of ferrochelatase, the enzyme required for conversion of protoporphyrin IX (PpIX) to heme, in tumors results in accumulation of PpIX in these tissues relative to the normal ones.[11] Such accumulation changes the relative concentration of these fluorophores thus altering the fluorescence spectra significantly, which in turn changes the peak heights of the emission bands of the two fluorophores. The scattering centers are known to enhance the fluorescence intensity.[31] Thus the large size of cell suspensions, higher density of cells and accumulation of more porphyrin in tumors may possibly contribute to the small peak at 630 nm wavelength region. Since all tissues studied in this work were non-necrotic and non-fungating, as corroborated histopathologically, protoporphyrin emission may not be due to bacterial action.[32]

A systematic analysis yields that, the normalized low pass powers at 4th level consistently differ between tissue types. Table-1 depicts normalized low pass powers of cancerous and corresponding normal tissues at the fourth level. The third column shows the values of fourth level normalized low pass powers and the second indicates the tissue type. It is observed from the table-1 that, the normalized low pass powers is more in cancerous tissues than its normal counterpart in ductal carcinoma (grade I, II, III), with a sensitivity of 80%. Interestingly, it is also found that, this value is less in tissues showing metastasis or with fatty tissues attached. These points need to be checked in a larger data base.

It is further observed that, normalized low pass powers of cancerous tissues decrease more slowly as a function of levels, as compared to their normal counter parts. Table-1 also gives the corresponding slope values, capturing these differences in the rate of fall; the fourth column shows the differences between the slope values. The slopes have been calculated by the linear fit of low pass power profile. It is observed that, fluctuations contribute significantly more to the total power in normal tissue fluorescence than their tumor counterparts. As Fig. 4(d) indicates, fluctuations contribute significantly to normal spectra and the average power measured through normalized low-pass coefficients decreases rapidly as a function of level for normal tissues. However for the cancerous cases, the average power remains quite high up to 6$^{th}$ level. It is worth reminding that due to Parseval's theorem, the total power is conserved in the orthogonal Haar basis. Therefore, if the fluctuations are high at certain level, the normalized low-pass power is small at the corresponding level.

**Table-I:** 4th level low pass powers of cancer and normal fluorescence spectra and slope values of the normalized six levels low pass power plots

| Sr. No. | Description of tissue type | 4th level low pass power of | | Slope values of the normalized six level low pass power | | |
|---|---|---|---|---|---|---|
| | | Cancer | Normal | Cancer | Normal | Difference |
| 1. | Ductal carcinoma grade-III | 0.907643 | 0.740373 | -0.011474 | -.037925 | -0.0265 |
| 2. | Cystosarcoma | 0.241762 | 0.161403 | -0.102240 | -0.108020 | -0.0058 |
| 3. | Ductal carcinoma grade-III | 0.581773 | 0.986615 | -0.012917 | -0.012797 | 0.0001 |
| 4. | Intraductal carcinoma (poorly defined) | 0.940750 | 0.881308 | -0.011890 | -0.016320 | -0.0044 |
| 5. | Ductal carcinoma grade-III | 0.934995 | 0.822425 | -0.102240 | -0.108020 | -0.0058 |
| 6. | Ductal carcinoma | 0.975588 | 0.953657 | -0.008624 | -0.016169 | -0.0075 |
| 7. | Cystosarcoma | 0.739220 | 0.884085 | -0.037675 | -0.020604 | 0.0171 |
| 8. | Ductal carcinoma grade-II positive for metastasis | 0.818529 | 0.903577 | -0.023613 | -0.015374 | 0.0082 |
| 9. | | 0.899315 | 0.887338 | -0.015041 | -0.017573 | -0.0025 |
| 10. | Ductal carcinoma grade-III positive for metastasis | 0.946701 | 0.922604 | -0.010214 | -0.011873 | -0.0017 |
| 11. | Ductal carcinoma grade-II | 0.717098 | 0.690419 | -0.031467 | -0.026810 | 0.0047 |
| 12. | Ductal carcinoma | 0.346617 | 0.672974 | -0.015834 | -0.025814 | -0.0100 |
| 13. | Ductal carcinoma | 0.995145 | 0.993628 | -0.005429 | -0.006476 | -0.0010 |
| 14. | Ductal carcinoma grade-II | 0.985754 | 0.967677 | -0.006719 | -0.015287 | -0.0086 |
| 15. | Ductal carcinoma grade-II | 0.988722 | 0.236050 | -0.005074 | -0.059195 | -0.0541 |
| 16. | Ductal carcinoma grade-III | 0.980755 | 0.961066 | -0.009140 | -0.019149 | -0.0100 |
| 17. | Ductal carcinoma grade-II | 0.994246 | 0.901727 | -0.007558 | -0.021904 | -0.0143 |
| 18. | Ductal carcinoma grade-III | 0.993581 | 0.985420 | -0.006816 | -0.092203 | -0.0854 |
| 19. | Ductal carcinoma grade-III positive for metastasis | 0.952731 | 0.956181 | -0.012288 | -0.011054 | 0.0012 |
| 20. | Ductal carcinoma grade-I | 0.988466 | 0.963549 | -0.006358 | -0.014993 | -0.0086 |
| 21. | Ductal carcinoma grade-II | 0.990160 | 0.915598 | -0.005952 | -0.024105 | -0.0182 |
| 22. | Cystosarcoma | 0.994139 | 0.918831 | -0.007244 | -0.016744 | -0.0072 |
| 23. | Lobular carcinoma | 0.305293 | 0.218660 | -0.064706 | -0.076012 | -0.0113 |

The fact that we have dealt with normalized energies, makes this comparison meaningful.

Similarly the slope, in the normalized low-pass coefficients (which are similar to the data itself since they represent averaged data points) indicates a sudden variation of the spectral profile in normal tissue fluorescence which is not present in cancerous cases.

Very interestingly, like the normalized low pass power, the cancerous samples with attached fatty tissues and the tissue types which are positive for metastasis show a higher steepness than the corresponding normal ones.

It may be noted that the above-mentioned three parameters also distinguish tumors of different grades. It is found that for grades I and II cancerous tissues, the values of the local low-pass maxima at the third quadrant are less than 0.2, but more than 0.2 in the grade III cancers, with a sensitivity of 75 %.

It is worth emphasizing that in the discrete wavelet transform, the average normalized low pass coefficients capture the broad band features in the spectra at a suitable scale and the normalized high pass coefficients extract the high frequency behavior. The orthogonal property ensures that the extracted features at different scales are independent of each other. It is clear that the normalized high pass coefficients can capture characteristic spectral variations, as also spectral fluctuations at different scales, whose property can be studied for tissue discrimination.

For Haar wavelets, the low pass coefficients represent average values of the data at different scales, whereas the high pass coefficients are the corresponding differences. At level-1 the wavelet coefficients therefore represents nearest neighbor differences which capture sudden changes in spectral intensities. Some of these variations can be ascribed to random fluctuations, arising from the randomization process in tissue light scattering and others to sharp changes in the spectral profile. The progressive averaging removes sharp variations, as also the random fluctuations. Hence, at higher levels the high pass coefficients capture the broader spectral variations and the low pass coefficients, the block averaged spectral intensity. The slopes of normalized low pass power shows sudden changes in spectral variations between 3rd and 4th levels. This discriminates the normal and diseased tissues well. It arises from the characteristic spectral variations present in the 3rd and 4th level, related to porphyrin emission.

In order to show the advantage of wavelet transform over the straightforward binning procedure (low pass), the high- pass features (since the normalized high pass coefficients are differences of intensities) have been carefully analyzed. The fact that, the normalized wavelet coefficients represent high-frequency variations makes them ideal to isolate characteristic spectral variations, from random ones. It is seen that high level normalized high pass coefficients do not give significant discrimination between tissue types (Table-2). However, the normalized spectral fluctuations, the normalized high-pass coefficients divided by their corresponding normalized low-pass coefficients, which are much more reliable for a statistical analysis at level-1 show clear distinction between diseased and normal tissues.

It is found that there is significant randomization of fluctuations[27] in case of cancerous tissue spectra, as compared to their normal counterparts. The fact that, cancerous tissues have nuclei of irregular shapes, which are much more tightly packed as compared to the normal tissues, may explain this observed feature. Hence in the difference of parallel and perpendicular spectra, which is free of the diffusive component, the spectral fluctuations in

**Table- II:** Result of the analysis of the high-power coefficients showing differences in grades of cancer and normal

| Sr. No. | | Description of tissue type | Values of high pass power of third level | | |
|---|---|---|---|---|---|
| | | | Cancer | Normal | Difference |
| | 1. | Ductal carcinoma grade-III | 2.18E-02 | 3.08E-02 | -8.99E-03 |
| | 2. | Cystosarcoma | 2.53E-02 | 3.85E-02 | -1.32E-02 |
| | 3. | Ductal carcinoma grade-III | 3.10E-02 | 4.71E-03 | 2.63E-02 |
| | 4. | Intraductal carcinoma (poorly defined) | 1.14E-02 | 3.12E-02 | -1.98E-02 |
| | 5. | Ductal carcinoma grade-III | 1.41E-02 | 2.27E-02 | -8.56E-03 |
| | 6. | Ductal carcinoma | 7.67E-03 | 1.70E-02 | -9.35E-03 |
| | 7. | Cystosarcoma | 6.37E-02 | 2.26E-02 | 4.11E-02 |
| | 8. | Ductal carcinoma grade-II positive for metastasis | 1.76E-02 | 2.63E-02 | -8.71E-03 |

cancer are devoid of characteristic fluctuations and show Gaussian random behavior[28]. The normal tissue's fluorescence spectra, on the other hand, retain their characteristic spectral variations, thereby showing a bigger spread in their wavelet coefficients and hence a larger standard deviation, as seen in Fig.6. A characteristic plot of the normalized high-pass power calculated for every scale is shown in Fig.7. For all scales, malignant tissues show

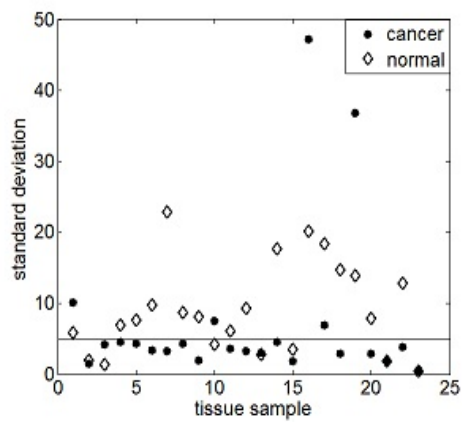

Fig.6. Standard deviations of the spectral fluctuations of the cancer and normal

tissues, captured through the high pass coefficients at level-1.

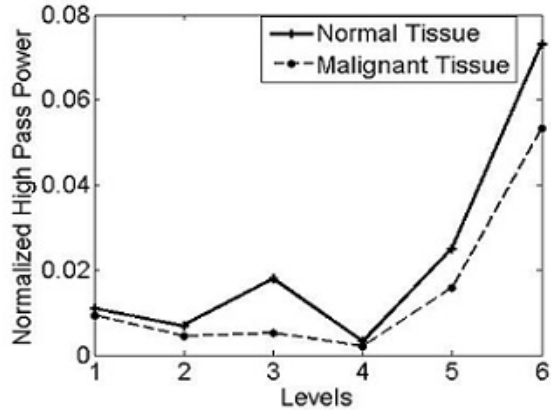

Fig.7. Normalized high pass power as a function of level, clearly low for cancer

for all levels and showing considerable difference at 3rd level.

weaker power compared to their normal counter-parts, the difference becomes considerably higher at 3rd level. Table-2 depicts the results of the analysis of the high-pass coefficients for all samples.

In conclusion, the systematic separation of variations at different wavelength scales from the broad spectral features pinpoints several quantifiable parameters to distinguish cancer and normal tissues. These distinguishable features are related with the biochemical and morphological changes. The spectral profile of diseased and the non-diseased tissues behave very differently, which manifest in the difference of the low pass power profiles. The fact that these characteristic signatures are based on higher level average coefficients makes them robust and less susceptible to experimental and statistical uncertainties. It is found that low level normalized high pass coefficients differ significantly between cancer and normal tissues. It is worth emphasizing that a straightforward averaging with arbitrary window sizes would not respect the orthogonality property, and hence would not allow an independent separation of fluctuations and average behavior at multiple scales. We observe that, scaling and translation are the key operations, which enable one to carry out a local analysis at desired scale and leads to independent wavelet coefficients, devoid of redundancy. This procedure is common to all discrete wavelets, where suitable form of weighted averaging and differentiation involving more than two points are carried out. The need for the early identification and constant monitoring of breast cancer for a large population makes this method eminently suitable since the same can be automated. We intend to analyze more samples and different type of cancers through both continuous and discrete wavelets to look for subtle spectral variations.